\documentclass[showpacs,showkeys,amsmath,amssymb,aps,twocolumn,prl]{revtex4-1}
\usepackage{graphicx}
\usepackage{dcolumn}
\usepackage{bm}
\begin{document}

\preprint{}

\title{Is subdiffusional transport slower than normal?}

\author{Igor Goychuk}
 \email{igor.goychuk@physik.uni-augsburg.de}
\affiliation{Institute of Physics, University of Augsburg, Universit\"{a}tstr. 1, D-86135 Augsburg, Germany}

\date{\today}

\begin{abstract} 
We consider anomalous non-Markovian transport of Brownian particles in
viscoelastic fluid-like media with  very large but finite macroscopic viscosity under the
influence of
a constant  force field $F$. The viscoelastic properties of the medium are characterized by a
power-law viscoelastic memory kernel which ultra slow decays in time on the time scale
$\tau$ of strong viscoelastic correlations.  The subdiffusive transport regime 
emerges transiently
for  $t<\tau$. However,  the transport becomes asymptotically normal for $t\gg\tau$.  
It is shown that even
though transiently the mean displacement and the variance both scale  sublinearly, i.e. 
anomalously slow,  in
time, $\langle \delta x(t)\rangle \propto F t^\alpha$, 
$\langle \delta x^2(t)\rangle \propto t^\alpha$,  $0<\alpha<1$,   
the mean displacement at
each instant of time  is nevertheless always larger than one obtained for normal transport in a 
purely viscous medium with the same macroscopic viscosity obtained in the 
Markovian  approximation. This can
have profound implications for the subdiffusive transport in biological cells as the notion of
``ultra-slowness'' can be misleading in the context of anomalous diffusion-limited transport and
reaction processes occurring on nano- and mesoscales.  
\end{abstract}

\pacs{05.40.-a, 05.10.Gg, 87.16.Uv}
\maketitle

\section{Introduction}

The widespread occurrence of anomalously slow diffusion and transport 
\cite{Shlesinger1,Scher,Bouchaud,Hughes,Shlesinger,Metzler,Balescu,new1,new2} 
in biological cells is still not commonly appreciated
in spite of a growing experimental evidence and 
support \cite{Wachsmuth,Qian,Seisenberg,Caspi,Tolic,Golding,Guigas,Weber,FNL3}. 
One of the main psychological obstacles on the way to a wider recognition is that ultra-slowness 
seems intuitively be rather
obstructive for the corresponding
diffusion-limited biochemical reactions, especially if it is caused by divergent mean residence times 
(MRTs) in trapping 
domains \cite{Shlesinger1,Scher,Hughes,Metzler} created by random meshwork of cell's cytoskeleton.
From this perspective, the occurrence of subdiffusion might be more associated with physics of dying, rather
than with physics of life. Moreover, the bulk of experimental biophysical data is traditionally 
interpreted in terms of
normal diffusion and even modern biophysics textbooks (see e.g. \cite{Jackson,Nelson}) discuss
only normal diffusion, despite appreciating  the fact of existence and importance of intracellular
molecular crowding which clearly obstructs diffusion via increase of the effective 
medium's viscosity \cite{Cell}. The increase of effective viscosity depends also on the size 
of diffusing particles. So, \emph{``for molecules smaller than 1 nm, it's similar to that of water;
for particles of diameter 6 nm (such as a protein of mass $10^5$ g/mol), it's about 3 times that
of water. For 50-500 nm particles, it's 30-300 times that of water"} \cite{Nelson} (p. 571). In
this respect, typical globular proteins are in the range of  2-20 nm (diameter) \cite{Nelson,Cell} and 
mRNA molecules are
about 400-800 nm in diameter \cite{Golding}. However, the traditional thinking and prejudges can
also be the reasons for overlooking  anomalous diffusion and transport regimes (probably mostly
transient) as recent experimental work uncovers \cite{Wachsmuth,Tolic,Golding,Guigas,Weber,Pan}.
Moreover, the occurrence of subdiffusion clearly depends not only on the size of single macromolecules,
but also on their concentration, i.e. on the degree of molecular crowding \cite{Guigas,Pan}.
 
One of approaches to anomalously slow  diffusion and transport is traditionally based on the 
assumption of divergent MRTs in trapping domains \cite{Shlesinger1,Scher,Hughes,Shlesinger,Metzler}. Of course,
MRT $\langle \tau\rangle$ in any finite spatial domain with linear size $\Delta x$
can never diverge in real life. However, it can  largely exceed 
a characteristic diffusion time, 
$\tau_D \sim (L^2/\kappa_\alpha)^{1/\alpha}$, required to 
subdiffusionally explore on average, $\langle \delta x^2(t)\rangle\propto \kappa_{\alpha} t^{\alpha}$ with $0<\alpha<1$, 
a \textit{finite} volume
with linear size $L\gg \Delta x$, where $\kappa_{\alpha}$ is the corresponding subdiffusion coefficient measured in 
${\rm cm^2}/{\rm sec}^{\alpha}$ \cite{GH04}. Then, the approximation of infinite MRTs becomes physically justified
on the relevant mesoscopic scale $L$.
 For very large times $t\gg \langle \tau\rangle $, the diffusion becomes normal, 
$\langle \delta x^2(t)\rangle\propto t$. However, the corresponding spatial scale can largely exceed $L$ and therefore 
the normal diffusion (and transport) regime can become of a little importance for certain processes
in mesoscopic biochemical reactors of living 
cells, such as e. g. passive transport of 
mRNA macromolecules or large 
globular proteins 
\cite{Golding,Guigas}, and in turn subdiffusion becomes of profound importance for such 
processes on mesoscopic scale. The approximation of divergent 
MRTs features the continuous time random walk (CTRW) approach to subdiffusion 
\cite{Shlesinger1,Scher,Hughes,Shlesinger,Metzler}. In this case, the position increments can be totally independent. 
Within the mean-field approximation, the CTRW transport is congruent \cite{Hughes} with jump-like 
transport in random potentials. Moreover, the overdamped continuous space Markovian Langevin dynamics in spatially
varying potentials
can be contracted onto such a semi-Markovian CTRW by doing properly 
spatial coarse-graining \cite{Reimann,Lindner}.
Then, a potential energy disorder can
result in anomalous diffusion and transport in agreement with the semi-Markovian CTRW theory as 
recent work nicely demonstrates 
\cite{Khoury} (see also paper by Lindenberg \textit{et al.} \cite{Lindenberg} in this Special Issue).     

Alternatively, subdiffusion
can result from the medium's viscoelasticity \cite{Qian,Guigas,Caspi,Mason,Amblard,Pan,G09,Szymanski}. 
It can be either
due to viscoelasticity of the polymer actin meshwork \cite{Guigas,Caspi,Mason,Amblard,Weigh}, or 
due to macromolecular crowding in complex fluids \cite{Weigh} as e.g. in cytoplasm of 
bacterial cells which are lacking static cytoskeleton \cite{Cell,Guigas,G09}. Statistical analysis 
of the experimental single particle diffusional trajectories in bacteria \cite{Golding} 
reveals  in fact the primarily
viscoelastic origin of subdiffusion \cite{Magdziarz}. A main result of Ref. \cite{Magdziarz} is that 
the fractional Brownian motion scenario (see below) is more likely
than one based on a semi-Markovian CTRW \cite{Remark}. The authors  of experimental work \cite{Szymanski}
came also to a similar conclusion.
Within this alternative subdiffusional scenario all the
moments of random time spent in finite spatial domains remain finite. The corresponding MRT is not only
finite but it scales down to zero with $\Delta x\to 0$. The physical reason for subdiffusion here is very 
different. It occurs due to long-time anticorrelations in the position increments \cite{G09}.
Considering Brownian particle of radius $R$, which starts to move at $t_0=0$ with velocity $\dot x(t)$ 
(we consider a one-dimensional 
case for simplicity), one expects it to experience a viscoelastic force
\begin{eqnarray}
F_{\rm v-el}(t)=-\int_0^t\eta(t-t')\dot x(t')dt',
\end{eqnarray}
where $\eta(t)$ is a frictional memory kernel whose Laplace-transform $\tilde \eta(s)$ is 
related to the frequency-dependent medium's viscosity $\tilde \zeta(i\omega)$ as
$\tilde \eta (s)=6\pi R \tilde \zeta(s)$. In the case of purely viscous fluids, and
in neglecting the hydrodynamic memory effects,
$\zeta(t)=2 \zeta_0\delta(t)$, where $\zeta_0$ is the fluid's macroscopic viscosity, so that $\eta(t)=2\eta_0\delta(t)$, where
$\eta_0=6\pi R\zeta_0$ is the Stokes viscous friction coefficient. 
For weakly viscoelastic fluids, $\zeta(t)=\zeta_0 \nu \exp(-\nu t)$, 
exponentially decays in time with rate $\nu$, and correspondingly $\eta(t)=6\pi R\zeta_0 \nu \exp(-\nu t)=\kappa \exp(-\nu t)$,
where $\kappa$ has dimension of a linear elastic force constant. This
corresponds to the Maxwell theory of viscoelasticity \cite{Maxwell} who derived the phenomenon of viscosity 
from medium's elasticity 
by assuming that the linear elastic 
force, $F_{\rm el}(t)=-\kappa [x(t)-x(0)]$, acting on the particle can relax in time with rate $\nu$, yielding a viscoelastic
force, i.e.
$\dot F_{\rm v-el}(t)=-\kappa \dot x(t)-\nu F_{\rm v-el}(t)$. 
Indeed, if the force relaxation is very fast with respect to the change of 
particle's velocity, then $F_{\rm v-el}(t)=-\int_0^t\eta(t-t')\dot x(t')dt'\approx -\eta_0\dot x(t)$, with $\eta_0=\kappa/\nu$, 
whereas in the opposite limit,
$F_{\rm v-el}(t)\approx -\kappa [x(t)-x(0)]$. For strongly viscoelastic media one expects that $\eta(t)$ decays in time
much slower than exponential and a power law decay $\eta(t)\propto t^{-\alpha}$ can serve as a better model. For a fluid-like
environment the effective macroscopic viscosity $\zeta_0=\int_0^{\infty}\zeta(t)dt$ should remain, however, finite.
It can be very large, but yet finite. Therefore, a long-time cutoff to the power-law must exist. 
In 1936, A. Gemant proposed a class of power-law viscoelastic models which are  consistent
with this demand \cite{Gemant,GReview}. Its particular representative corresponds in the Laplace space to
\begin{eqnarray}\label{Gemant1}
\tilde \eta(s)=\int_0^{\infty}\eta(t)\exp(-st)dt=\frac{\eta}{1+(s\tau)^{1-\alpha}},
\end{eqnarray}     
 in our notations. Here, $\tilde \eta(0)=\eta$ is an effective asymptotic 
 friction coefficient and $\tau$ presents a long-time
 memory cutoff. The corresponding memory kernel is approximately
 \begin{eqnarray}\label{Gemant}
 \eta(t)\approx \frac{\eta_{\alpha}}{\Gamma(1-\alpha) t^\alpha}
 \end{eqnarray}
 for $t\ll \tau$, where $\eta_{\alpha}=\eta\tau^{\alpha-1}$, and $\Gamma(x)$ is
the familiar gamma-function. For $t\gg \tau$, $\eta(t)$ decays also in accordance with a power law,
$\eta(t)\propto t^{\alpha-2}$, i.e. elastic correlations are still rather strong. However, the corresponding
integral converges ensuring that the asymptotic friction coefficient $\eta$ is finite.
 In the limit of infinitely large medium's viscosity yielding $\eta\to\infty$, and infinitely long memory range, $\tau\to\infty$, 
 with $\eta_{\alpha}=\eta\tau^{\alpha-1}$ kept constant, Eq. (\ref{Gemant}) becomes exact,  
 $\tilde \eta(s)=\eta_{\alpha}s^{\alpha-1}$, and
\begin{eqnarray}\label{fract1}
F_{\rm v-el}(t) & = &-\int_0^t\frac{\eta_{\alpha}}{\Gamma(1-\alpha)(t-t')^\alpha}\dot x(t')dt' \nonumber \\
& := &-\eta_{\alpha}
 \sideset{_{0}}{_{*}}{\mathop{D}^{\alpha}} x(t)\;,
\end{eqnarray}   
where the last equality defines fractional Caputo derivative of the order $0<\alpha<1$ \cite{Gorenflo} acting 
on $x(t)$. $\eta_{\alpha}$
can be named the fractional friction coefficient.
Clearly, Eqs. (\ref{Gemant},\ref{fract1}) can serve as a good approximation only for the times $t\ll \tau$. In the focus
of this Letter is but the entire time evolution, interpolating between transient subdiffusion and asymptotically
normal diffusion behavior. For example, $\tau$ can correspond to the time scale of seconds or  minutes, 
and then subdiffusion emerges
on the time scale from microseconds to seconds or minutes, as in biological cells \cite{Golding,Guigas,Weber,Pan}. 

\section{Simple model}

We continue with a non-Markovian generalized Langevin equation (GLE) description \cite{Kubo1,Zwanzig1,Kubo2} 
for an overdamped Brownian particle neglecting inertial effects. Then,
\begin{eqnarray}\label{GLE}
\int_0^t\eta(t-t')\dot x(t')dt'=f(x,t)+\xi(t),
\end{eqnarray}
where $f(x,t)$ is a generally nonlinear force acting on the particle and $\xi(t)$ is a thermal random force of 
the environment.
It is Gaussian, unbiased on average, and obeying the fluctuation-dissipation relation,
\begin{eqnarray}\label{FDR}
\langle \xi(t)\xi(t')\rangle = k_B T \eta(|t-t'|),
\end{eqnarray}
at the environmental temperature $T$.
This is required for the consistency with thermodynamics at thermal equilibrium. In the above-mentioned 
limit $\tau\to\infty$, $\eta\to\infty$, $\eta_{\alpha}=const$, GLE (\ref{GLE}) is named also the 
fractional Langevin equation \cite{new2} upon 
the use of the corresponding abbreviation (\ref{fract1}) for its lhs. In this limit, $\xi(t)$ is nothing
else but the fractional Gaussian noise by Mandelbrot and van Ness \cite{Mandelbrot} which presents an instance
of $1/f$ noise with the spectral power density $S(\omega)\propto 1/\omega^{1-\alpha}$. 
Notice that generally the lower integration limit in Eq. (\ref{GLE}) 
is $t_0\to-\infty$. It can be replaced, however, with $t_0=0$ since
we assume that the particle starts to move at this time being initially localized, 
i.e. $v(t)=\dot x(t)=0$ for $t<t_0$.

\begin{figure}
 \includegraphics[width=0.7\columnwidth]{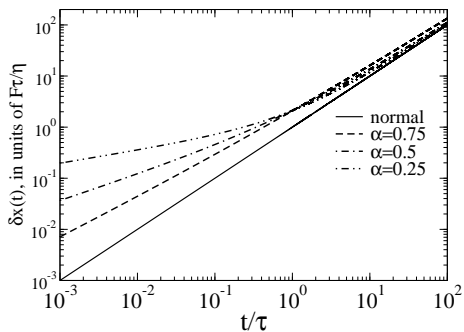}
 \caption{Mean displacement (in units of $F\tau/\eta$) versus time (in units of $\tau$) 
under the influence of constant force $F$ 
for several different 
values of $\alpha$ and the same $\eta$, $\tau$.
The limit of normal diffusion is achieved asymptotically from above. Transiently subdiffusing particles always cover larger
distances than normally diffusing particles with the same asymptotic frictional force constant $\eta$.
 \label{Fig1}} 
\end{figure}

Let us consider the transport under the influence of  constant
force $F$. Then, the above GLE can be easily solved formally using the Laplace-transform method, for any memory 
kernel $\eta(t)$. Transforming back to the time domain for the memory kernel (\ref{Gemant1}),
 one obtains for the averaged mean displacement the simple result,
\begin{eqnarray}\label{solution1}
\langle \delta x(t)\rangle & = &\frac{Ft}{\eta}+\frac{Ft^{\alpha}}{\eta_{\alpha}\Gamma(1+\alpha)}\nonumber \\
 & = &\frac{F\tau}{\eta}
\left [\frac{t}{\tau}+ \frac{1}{\Gamma(1+\alpha)}\left (\frac{t}{\tau}\right )^{\alpha}\right ] \;.
\end{eqnarray}
This exact solution is compared in Fig.\ref{Fig1} with the solution of the 
ordinary Langevin equation with memoryless friction
for the particle which experiences the same frictional force for the whole time span as our particle asymptotically,
or, said differently, the result of the Markovian approximation to the considered dynamics. 
Clearly, for all times our particle moves in fact faster, covering larger distances and approaching gradually the limit 
of normal diffusion from above. The instant time-dependent ensemble-averaged velocity 
$\langle v(t)\rangle:=\langle \delta \dot x(t)\rangle=F/\eta+F/[\eta_{\alpha} \Gamma(\alpha)t^{1-\alpha}] $ 
is also always larger
than its asymptotic value $F/\eta$ (a spurious singularity at $t=0$ can be eliminated, if to take
the initial inertial effects into account). 
Mathematically, this is simply because $x^\alpha \gg x$, for $x\ll 1$ and $0<\alpha<1$. 
The physics is also  clear. In the limit $\eta\propto \zeta_0\to\infty$ the normally diffusing particle
is get localized, $\langle \delta x(t)\rangle \to 0$, whereas our particle still moves, 
but ultra slow (per definition) since
$\langle \delta x(t)\rangle \propto (t/\tau)^\alpha$. Therefore, the ``ultra-slow'' moving particle can cover larger distances.
Furthermore, for any memory kernel in the studied model and for arbitrary constant $F$ 
the variance of the particle position, $\langle \delta x^2(t)\rangle=
\langle x^2(t)\rangle-\langle  x(t)\rangle^2$ obeys \cite{Kubo1}
\begin{eqnarray}\label{var1}
\langle \delta x^2(t)\rangle=\frac{2 k_BT}{F}\langle \delta x(t)\rangle,
\end{eqnarray} 
and therefore it follows to the same pattern as in Eq. (\ref{solution1}) and Fig. \ref{Fig1},
\begin{eqnarray}\label{solution2}
\langle \delta x^2(t)\rangle =2\kappa_1 t+2\kappa_{\alpha}t^{\alpha}/\Gamma(1+\alpha)\;.
\end{eqnarray}
Here, $\kappa_{\alpha}$ is fractional diffusion coefficient related to temperature and fractional 
friction coefficient
by the generalized Einstein-Stokes relation, $\kappa_{\alpha}=k_BT/\eta_{\alpha}$, which
contains the standard one, $\kappa_{1}=k_BT/\eta$, as a particular case.

We suppose that our observation is rather general. For example, 
the results in Ref. \cite{Golding} seem to agree with our line of reasoning. Indeed, mRNA macromolecules 
have in the related experiments radii in the range of 200-500 nm. Furthermore, the normal diffusion coefficient in
water was found to be $\kappa_1= 1\;\mu{\rm m^2/sec}$ (see Supplementary Material in \cite{Golding}). From this,
given the water viscosity $\zeta_w=0.9 \cdot 10^{-3}$ Pa$\cdot$sec, one can estimate the corresponding radius as 
$R=k_BT/(6\pi\zeta_w\kappa_1)$ which gives $R\approx 242$ nm for $T=300$ K. Let us assume that $R\approx 250$ nm. 
Then, the corresponding macroscopic normal diffusion coefficient in cytosol should be by the factor 
of $r\approx 300$ smaller than one in water \cite{Nelson}
(see the above quotation in Introduction). This yields 
$\kappa_1^{(cyt)}\approx\kappa_1/r\approx 3.33\cdot 10^{-3}\;\mu{\rm m^2/sec}$.  However, the experiment yields
not normal but 
subdiffusion with $\alpha\approx 0.7$ (see Fig. 2(a) in \cite{Golding}) and $\kappa_{\alpha}$ 
in the range from $10^{-3}$ to $10^{-2}$ $\mu{\rm m^2/sec^{0.7}}$ \cite{Remark}. 
Assuming $\kappa_{\alpha}=10^{-2}\;\mu{\rm m^2/sec^{0.7}}$ for this value of 
$R$ (smaller particles in experiment should also 
subdiffuse faster) one can
conclude that subdiffusion can indeed cover
larger distances than normal diffusion with $\kappa_1^{(cyt)}\sim\kappa_1/r$. Furthermore, one can estimate the transition
time $\tau$. Given
the relation $\tau=(\eta/\eta_{\alpha})^{1/(1-\alpha)}=(\kappa_{\alpha}/\kappa_1^{(cyt)})^{1/(1-\alpha)}$ which follows 
within our model one obtains for it  $\tau\approx 55 $ sec. This is a rather reasonable estimate since 
subdiffusion regime lasts in those experiments up to 30 sec, cf. Fig. 2(a) in \cite{Golding}.

\section{Conclusions}

The discussed phenomenon might seem paradoxical, even though its explanation is almost trivial. 
Nevertheless, it has 
profound implications for subdiffusion in  biological cells.
First of all, the occurrence of subdiffusion on some transient time scale $\tau$ and the corresponding
mesoscopic spatial scale $L\sim (2\kappa_{\alpha}\tau^{\alpha})^{1/2}$ 
does not contradict to the bulk of macroscopic experimental data
indicating typically a normal diffusion \cite{Nelson}. 
Even more important, the overall transport is in fact faster than
its long time normal asymptotics that results from a drastic reduction of the 
effective normal diffusion coefficient for large macromolecules
due to molecular crowding effects in cytosol. Very important is also the fact that on the time scale 
$t\ll\tau$, by use of the memory kernel (\ref{fract1}) as approximation, the resulting random process is nothing
else the fractional Brownian motion (fBm) \cite{Mandelbrot,G09}. The fractal dimensionality of the 
fBm trajectories is
known to be $d_f=2/\alpha$ \cite{Feder}, i.e. $d_f=2$ for the normal Brownian motion.  However, 
for $\alpha\leq 2/3$ and for  3d Euclidean 
embedding dimension 
it becomes $d_f=3$ (a fractal trajectory cannot have larger Hausdorff dimension that the Euclidean space
in which it is living). This means that a subdiffusionally searching particle explores much more 
thoroughly  finite volumes than a normally diffusing particle on the same spatial and time scales.
Therefore, transient subdiffusion can provide only advantages for the diffusion-limited reactions in tiny biochemical reactors
of living cells, which are densely stuffed with different macromolecules, 
without some principal drawbacks. This is rather unexpected and paradoxical conclusion.

\section*{Acknowledgments}

This paper is dedicated to Professor P.V.E. McClintock on occasion of his 70th birthday. 
Support of this work by the Deutsche Forschungsgemeinschaft, grant GO 2052/1-1
is gratefully acknowledged.


\begin{thebibliography}{0}
\bibitem{Remark}Visible scattering in the distribution 
of single trajectories can reflect e.g. distribution of mRNA molecules over their
sizes, binding properties, etc. \cite{Cell}; this being in a combination with the absence of self-averaging even 
in the case of ergodic diffusion for experimental trajectories of finite length.
\end{thebibliography}
\end{document}